\newcommand{\aap}{Astronom. Astrophys.}
\newcommand{\icarus}{Icarus}
\newcommand{\psj}{PSJ}
\newcommand{\pasp}{PASP}
\newcommand{\ssr}{Space~Sci.~Rev.}
\newcommand{\aj}{AJ}
\newcommand{\apj}{ApJ}

\documentclass{article}

\usepackage[english]{babel}

\usepackage[letterpaper,top=2cm,bottom=2cm,left=3cm,right=3cm,marginparwidth=1.75cm]{geometry}

\usepackage{amsmath}
\usepackage{graphicx}
\usepackage[colorlinks=true, allcolors=blue]{hyperref}
\usepackage{lipsum}
\usepackage{gensymb}
\usepackage{multirow}
\usepackage{lineno}
\usepackage{natbib}
\usepackage{booktabs}

\title{Refined rotational state and shape model of (98943) Torifune ahead of the Hayabusa2\# flyby}

\author{P. Fatka\textsuperscript{1*},
        P. Pravec\textsuperscript{1},
        P. Scheirich\textsuperscript{1},
        P. Ku\v{s}nir\'{a}k\textsuperscript{1},
        K. Hornoch\textsuperscript{1},
        H. Ku\v{c}\'akov\'a\textsuperscript{1},\and
        M. Hirabayashi\textsuperscript{2}\\[1ex]
        \textsuperscript{1}Astronomical Institute of the Czech Academy of Sciences, Fri\v{c}ova 298, Ond\v{r}ejov,\\ CZ-25165, Czech Republic\\
        \textsuperscript{2}Georgia Institute of Technology, 620 Cherry Street, Atlanta, GA, 30332, USA\\
\texttt{$^*$petr.fatka@asu.cas.cz}
}
\begin{document}

\maketitle

\begin{abstract}
The Japan Aerospace Exploration Agency (JAXA) Hayabusa2\# mission will perform a high-speed flyby of near-Earth asteroid (98943) Torifune on 5 July 2026, offering a rare opportunity to compare a pre-encounter spin and shape model derived from convex light-curve inversion with spacecraft imaging. We refined Torifune’s rotational state and convex shape model using previously published dense light curves, new photometry data obtained in late 2025, and selected sparse photometry data from the Asteroid Terrestrial-impact Last Alert System (ATLAS). A key methodological contribution is a per-measurement weighting scheme that accounts for heterogeneous data quality and cadence through light-curve scatter, rotational-phase-dependent brightness, and correlation-time downweighting of closely spaced measurements. We also estimated uncertainties with an apparition-constrained block bootstrap, resampling complete light curves while preserving the dense/sparse composition of the dataset and retaining coverage of each apparition. This procedure provides realistic uncertainties for the determined sidereal period, pole direction, and shape parameters. We refine the sidereal period to $P_{\mathrm{sid}} = 5.0215221^{+0.0000011}_{-0.0000007}$ h and confirm a prograde spin state with the axis near the north ecliptic pole. The nominal pole solution is $(\lambda, \beta) = (314\degree, +84\degree)$, with a $5.4\degree$ angular uncertainty. The dynamically equivalent ellipsoid has $a/b = 1.66^{+0.03}_{-0.07}$, whereas its polar axis remains less well constrained, with $b/c = 1.49^{+0.44}_{-0.29}$. All quoted uncertainties correspond to the two-sided 95\% intervals of the block-bootstrap distributions. The forthcoming Hayabusa2\# flyby should enable direct evaluation of which elements of this pre-encounter convex inversion model are robust, particularly the pole orientation and global silhouette, and may help constrain the remaining uncertainty in the polar dimension of the body.\end{abstract}
\section{Introduction}
Asteroid (98943) Torifune (formerly 2001 CC$_{21}$) is a near-Earth asteroid of particular interest because it is the target of the Hayabusa2 extended mission, Hayabusa2\# \citep[][]{Hirabayashi2026, Tsuda2013,Watanabe2017}, with a flyby expected on July 5, 2026. Torifune will be the first asteroid encountered by Hayabusa2\#, prior to its later rendezvous with 1998 KY$_{26}$ in 2031. The upcoming encounter makes Torifune an especially timely target for ground-based physical characterization, as constraints on its rotational state and shape are directly relevant to maximizing the scientific return of the flyby. In particular, an improved spin and shape model can help predict the viewing geometry of the encounter and support observation planning aimed at imaging as much of the surface as possible during the short, high-velocity passage.

Torifune's spin state and shape were previously modelled by \citet[][]{Fatka2025, Popescu2025} using the photometric dataset available at that time. The new observations presented here improve the temporal coverage and the sampling of observing geometries, including three newly covered apparitions. In order to refine the sidereal period, spin-axis orientation and convex shape model prior to the spacecraft encounter, we also employ a new data-weighting strategy in the modelling and uncertainty estimation using a block bootstrap method.

Beyond its immediate practical value for the Hayabusa2\# flyby, Torifune also provides a rare opportunity to test the performance of the convex light-curve inversion method. For most asteroids, shape models derived from disk-integrated photometry cannot be compared directly against images taken by a spacecraft. In the case of Torifune, however, the forthcoming flyby images should enable an unusually direct comparison between a pre-encounter convex shape model and the actual figure of the asteroid. Torifune therefore represents an important benchmark for evaluating how well convex inversion recovers the global shape and spin properties of near-Earth asteroids under realistic data limitations.

\section{Dataset}\label{sec.dataset}
We used all photometric data of asteroid (98943) Torifune previously published by \citet{Fatka2025}, spanning 2002 to 2023, and complemented them with newly available observations. In particular, we added the dataset provided by \citet{Fornasier2024}, which covers the early 2024 apparition not included in the previous study, together with selected light curves from the Solar System Catalog V3 of the Asteroid Terrestrial-impact Last Alert System \citep[ATLAS;][]{Tonry2018}, which newly sample the late 2024 apparition. We adopted the same selection criteria for the ATLAS data as \citet{Fatka2025}. We retained only nightly sessions with at least three measurements, allowing consistency checks, and with reported errors of $\leq 0.05$ mag.

We also obtained new photometric observations of Torifune with the Danish 1.54-m telescope\footnote{For the telescope and instrument description, see \citet{Fatka2025}.} at La Silla Observatory, Chile. In total, we carried out seven observing sessions between 23 October and 20 December 2025, covering phase angles from $20\degree$ to $46\degree$. Detailed observing circumstances are listed in Table~\ref{tab.new_obs}. A summary of the full dataset is listed in Table~\ref{tab.obs_all}.

\begin{figure}[ht!]
   \centering
    \includegraphics[width=0.8\hsize]{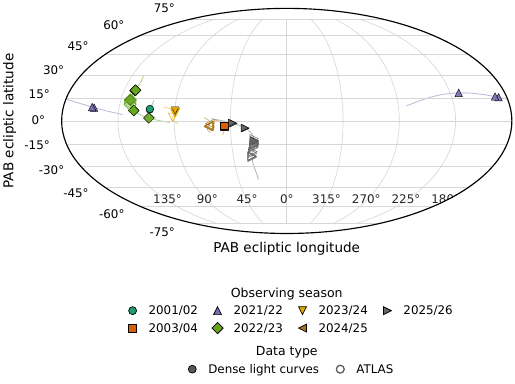}
    \caption{Phase-angle-bisector coverage of the Torifune observations in ecliptic coordinates. Symbols mark the PAB longitude and latitude at the mid-time of each observing session, using the JPL Horizons ICRF/J2000 PAB convention. Marker shapes and colors distinguish observing seasons, many of which span a calendar-year boundary. Filled symbols denote dense light curves, whereas open symbols denote sparse ATLAS observations. Faint solid curves trace the PAB evolution for each observed season from one month before the first observation to one month after the last observation. Ecliptic longitude is plotted in the astronomical sense, increasing to the left.}
     \label{fig.pab_dist}
\end{figure}

The combined dataset extends the temporal baseline and the range of observing geometries relative to the dataset used by \cite{Fatka2025}. As shown in Fig.~\ref{fig.pab_dist} the 2025–2026 observations extend the phase-angle-bisector coverage in ecliptic longitude and, more importantly, in ecliptic latitude, thereby improving sensitivity to the polar dimension of the shape model. In particular, the inclusion of three newly sampled apparitions strengthens the constraints on the updated spin and shape model by extending both the temporal baseline and the range of observing geometries.

\begin{table}[ht!]
\caption{\label{tab.new_obs} Circumstances of the new observations of asteroid (98943) Torifune.}
\centering
\begin{tabular}{lccc}
\toprule
Date (UT) & Exp. time (s) & Number of points & $\alpha$ ($^\circ$) \\
\midrule
2025-Oct-23.0   & ~45           & 35            & 21 \\
2025-Oct-24.1   & ~45           & 40            & 20 \\
2025-Oct-25.3   & ~45           & 26            & 20 \\
2025-Nov-26.1   & 150           & 12            & 38 \\
2025-Nov-27.2   & 150           & ~7            & 38 \\
2025-Dec-18.1   & 130           & 17            & 46 \\
2025-Dec-20.1   & 120           & 20            & 46 \\
\bottomrule
\end{tabular}

\par\vspace{0.5em}
\begin{minipage}{0.48\textwidth}
\raggedright
\footnotesize
Note: The date corresponds to the UT mid-time of each observing session, and $\alpha$ is the solar phase angle. All observations were obtained with the Danish 1.54-m telescope at La Silla Observatory in the standard Cousins $R$ filter.
\end{minipage}
\end{table}

\section{Light-curve inversion}

We determined the spin state and convex shape model using the standard light-curve inversion method developed by \citet{Kaasalainen2001a,Kaasalainen2001b}. The computations were performed with the C implementation of the inversion code rewritten by Josef \v{D}urech. We further modified this version to allow individual photometric measurements to be assigned different weights, so as to account for differences in photometric quality, reduction procedures, and possible systematics among the datasets used in this work.

\subsection{Weighting scheme}

The photometric datasets used in this work are heterogeneous, and the reported measurement uncertainties are either derived using different reduction procedures or are not available at all. In addition, the quoted uncertainties are often only formal and may underestimate the true scatter of the data. For this reason, we did not use the reported uncertainties directly, but instead estimated the weights of individual measurements using a three-step procedure.

First, for each dense light curve, we fitted a Fourier series up to the highest statistically significant order, determined by an F--test, and calculated the root mean square (RMS) of the residuals. All points within a given dense light curve were then assigned the same basic weight, computed as
\begin{equation}
w_{\mathrm{lc}} = 1/ \sqrt{RMS^2 + 0.02^2},
\end{equation}
where the constant floor term of 0.02 represents the uncertainty associated with the model simplification that is seen in light-curve inversion models \citep[][]{Muinonen2022, Durech2023}. 

Sparse ATLAS data cannot be approximated reliably by a Fourier series because of the low number of points obtained per night. Therefore, in the initial step, we used nightly averaged reported uncertainties of the ATLAS measurements as a proxy for the RMS and computed $w_{\mathrm{lc}}$ in the same way. After obtaining the best-fit solution, we calculated a scaling factor
\begin{equation}
s = \sqrt{\frac{1}{N_{\mathrm{ATLAS}}} \sum_{i \in \mathrm{ATLAS}} \left(\frac{r_i}{\sigma_i}\right)^2}
\end{equation}
separately for the two ATLAS filters, where $r_i$ denotes the residuals and $\sigma_i$ the corresponding reported uncertainties. We obtained $s_{\mathrm{c}} = 1.29$ for the \textit{cyan} filter, indicating that the reported uncertainties are underestimated by 29 \%, and $s_{\mathrm{o}} = 0.89$ for the \textit{orange} filter, indicating a slight overestimation. After correcting the ATLAS uncertainties accordingly, we repeated the full procedure of period search and spin and shape determination. We note that these scaling factors were derived only from the subset of ATLAS data satisfying the selection criteria mentioned in Sect.~\ref{sec.dataset}, and not from the full ATLAS dataset available for Torifune.

Second, we accounted for the fact that closely spaced measurements are not fully independent, as they may be affected by temporary systematic errors, for example when the asteroid passes close to a background star. Following the approach of \citet{Scheirich2021}, we defined a correlation time $d$ and calculated how many other data points, $K_i$, are within $\pm d/2$ from the given point. We then assigned a weight $w_{\mathrm{corr},i} = 1/K_i$ to that data point. This approach effectively reduces the number of independent measurements and prevents the inversion from being driven disproportionately by high-cadence light curves at the expense of more sparsely sampled data. We empirically set $d$ as 1/60 of the rotational period of Torifune (about 5 minutes), reducing the number of independent measurements by 41 \%. We also tested values 1/30, 1/45, and 1/90 of the rotational period for $d$, and found only subtle differences in the best-fit model. The initial weights used in the modelling were then defined as
\begin{equation}
w_{\mathrm{ini}} = w_{\mathrm{lc}} \cdot w_{\mathrm{corr}}.
\end{equation}

Finally, once the best-fit solution had been found, we were able to determine the rotational phase of the individual measurements more accurately. We then used the normalized brightness values, centred around unity, to define an additional weight factor, $w_{\mathrm{phase}}$, which accounts for phase-dependent variations in data quality. This reflects the expected increase in signal-to-noise ratio with brightness under the assumption of photon-noise-dominated uncertainties. The final weights were thus defined as

\begin{equation}
w_{\mathrm{fin}} = w_{\mathrm{lc}} \cdot w_{\mathrm{corr}} \cdot w_{\mathrm{phase}},
\end{equation}
and the full procedure of period search and spin and shape modelling was repeated using these weights. We note that incorporating $w_{\mathrm{phase}}$ into the final weights produces a minor change (below the estimated uncertainties) in the best-fit model.

\subsection{Uncertainty estimates}
The uncertainties of the sidereal period, spin-axis direction, and shape parameters were estimated by means of a block bootstrap. This approach is well suited to light-curve inversion, because the observations are naturally grouped into individual light curves and measurements within a given group cannot be considered statistically independent. Moreover, dense and sparse-in-time photometry provide complementary constraints on the inversion and should therefore be resampled in a way that preserves their relative contribution to the model \citep{Hanus2013}. 

In each bootstrap realisation, the resampling unit was one complete light curve. The resampled dataset was generated by random sampling with replacement from these blocks, while preserving the original number of dense and sparse light curves. In addition, to preserve the coverage of observing geometries across apparitions, we required that each bootstrap realisation include at least one light curve from each apparition represented in the original dataset. All measurements belonging to a selected block were retained together with their original sampling and weights. Such block-wise resampling preserves the internal structure of the observations and is more appropriate for clustered data than resampling individual measurements independently \citep{Lahiri2003, Field2007}. A total of 2000 realisations of such resampling were conducted.

For each resampled dataset, we repeated the full light-curve inversion and stored the corresponding best-fit solution. We report central percentile intervals rather than symmetric standard-deviation errors, because the bootstrap distributions are not necessarily Gaussian and may be asymmetric. The 68\% interval (16th--84th percentiles) is quoted as the closest bootstrap analogue of a $1\sigma$ uncertainty, while the 95\% interval (2.5th--97.5th percentiles) is given as a more conservative measure of solution stability.

\section{Results}
\begin{figure}[ht!]
   \centering
    \includegraphics[width=0.5\hsize]{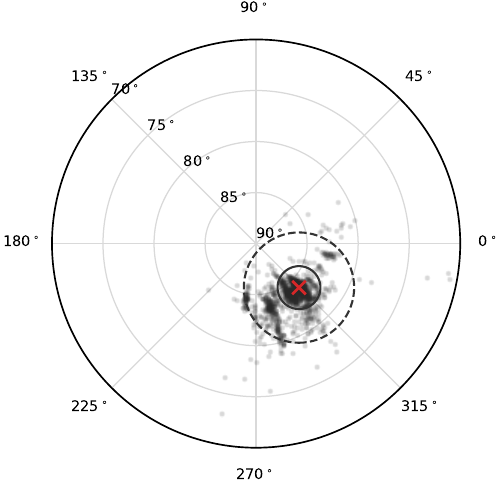}
    \caption{Polar projection of the northern ecliptic hemisphere showing the distribution of spin-axis directions from the bootstrap realisations (grey points). The red cross marks the nominal best-fit pole. The black solid and dashed circles, centred on the nominal solution, correspond to the 68\% and 95\% bootstrap percentile regions, respectively.}
     \label{fig.poles}
\end{figure}

Our modelling efforts resulted in a unique and tightly constrained rotational period $P_{\mathrm{sid}} = 5.0215221^{+0.0000011}_{-0.0000007}$ h as shown in Fig.~\ref{fig.periods}. This value agrees with \citet{Fatka2025, Popescu2025} and is more precise due to the longer time span and improved geometry coverage. The spin-axis direction is also unique and robust. The best-fit pole directions of the bootstrap realisations form a compact cluster around the nominal spin axis, see Fig.~\ref{fig.poles}. The nominal solution is at ecliptic coordinates (J2000 epoch) $(\lambda,\beta) = (314\degree, +84\degree)$ with a 95th-percentile angular uncertainty of 5.4\degree, which agrees with our previous findings in \citet{Fatka2025}. The uncertainty region is roughly half the size of that reported in \citet{Fatka2025}. However, we note that the uncertainty-estimation methods differ.A comparison overview of derived Torifune models is shown in Table~\ref{tab.res_comp}. The comparison shows that all published models agree closely in sidereal period and place the pole near the north ecliptic pole, whereas the inferred shape parameters, especially the polar flattening, remain model-dependent. The large differences in $b/c$ among the available convex solutions therefore identify the polar dimension as a key quantity to be tested by the flyby. In this sense, the Hayabusa2\# encounter will not merely refine the physical properties of Torifune, but will also determine which aspects of the pre-encounter light-curve inversion are genuinely constrained by the existing data. Although the pole longitudes in Table~\ref{tab.res_comp} differ, all solutions lie close to the north ecliptic pole, so the actual angular separations between them are small. Because the pole lies close to the north ecliptic pole, the reported form of the pole uncertainty is more meaningful than uncertainties of longitude and latitude reported separately.

\begin{figure}[ht!]
   \centering
    \includegraphics[width=0.8\hsize]{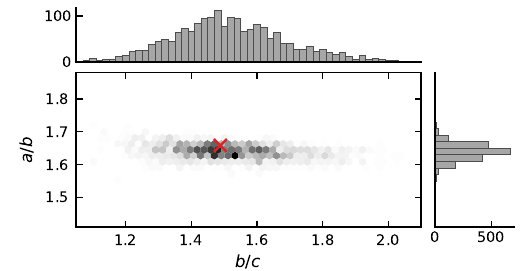}
    \caption{Axis ratios of the dynamically equivalent ellipsoids corresponding to the bootstrap best-fit shape models. The hexagonal shading shows the local density of bootstrap solutions in the axis-ratio plane, with darker regions indicating higher densities. The bin width in both histograms is 0.02.}
     \label{fig.axis_ratio}
\end{figure}

\begin{figure}[ht!]
   \centering
    \includegraphics[width=0.8\hsize]{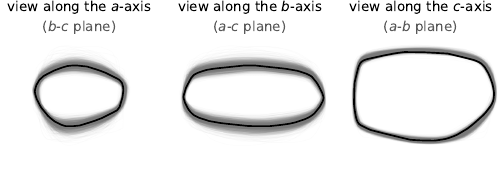}
    \caption{Silhouette uncertainty of the nominal shape model derived from bootstrap solutions. The black curve shows the nominal model, while the grey contours show the silhouettes of the bootstrap shape realisations. The three panels present orthographic projections viewed along the $a-, b-,$ and $c-$axes of the nominal model. All projections are shown at the same scale. The bootstrap models were scaled to match the nominal model along the a-axis.}
     \label{fig.shape_unc}
\end{figure}

\begin{figure}[ht!]
   \centering
    \includegraphics[width=0.9\hsize]{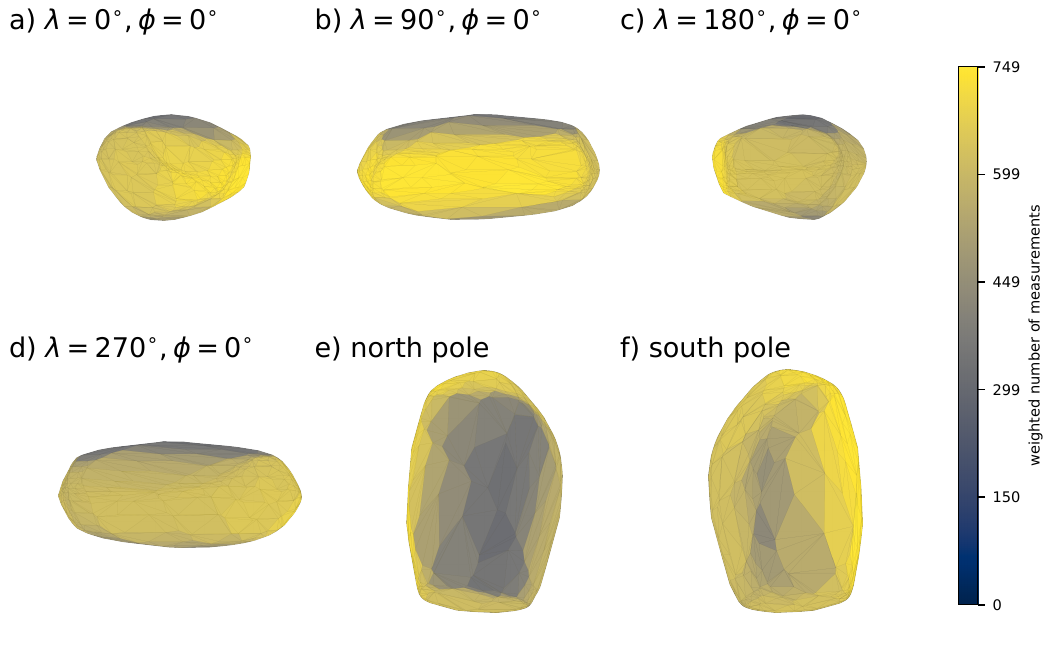}
    \caption{Cumulative observational coverage of the nominal convex shape model. Facets are colored by the number of selected photometric measurements for which the facet is simultaneously visible from the observer and illuminated by the Sun. A facet contributes when both the emission-angle and incidence-angle cosines are positive in the body-fixed frame. Panels show views toward body-fixed longitudes ($0^\circ$), ($90^\circ$), ($180^\circ$), and ($270^\circ$), followed by north-pole and south-pole views. The color bar gives the weighted number of contributing measurements.}
     \label{fig.observed_parts_all}
\end{figure}

The shape (see Fig.~\ref{fig.best_shape}) and its parameters are less tightly constrained than the spin state, as expected for convex inversion from disk-integrated photometry. The $a/b$ axis ratio of the corresponding dynamically equivalent ellipsoid is relatively well constrained with $a/b = 1.66^{+0.03}_{-0.07}$, while the $b/c$ ratio is far more loosely constrained with $b/c = 1.49^{+0.44}_{-0.29}$ as shown in Fig.~\ref{fig.axis_ratio}. This interpretation is consistent with the PAB distribution in Fig.~\ref{fig.pab_dist}, which, despite the improved 2025–2026 coverage, remains much more extended in longitude than in latitude. A complementary view of this result is provided by Fig.~\ref{fig.observed_parts_all}, which shows that the nominal model is sampled highly non-uniformly by the available observations, especially in the polar views. The silhouettes of the nominal shape together with those of the bootstrap realisations are shown in  Fig.~\ref{fig.shape_unc} illustrating different uncertainties along each principal axis. Reported uncertainties in this Section correspond to the 2.5th and 97.5th percentiles of the 2000 bootstrap realisations. For $1\sigma$--like uncertainty estimates, see Table~\ref{tab.results}.

\begin{table*}[tb]
\caption{\label{tab.res_comp}Comparison of published convex-model parameters for asteroid (98943) Torifune.}
\begin{tabular}{lrrrrr}
\toprule
Source              & $P_{sid}$ (h) &  $\lambda$ (\degree) & $\beta$ (\degree)& $a/b$    & $b/c$ \\
\midrule
\cite{Popescu2025}  & 5.021516       &  301         & +89           & 2.47                   & 1.06 \\
75\% random subsampling & $\pm 0.000106$ & \multicolumn{2}{c}{radius $\approx6.2$ \degree}  & $^{+0.96}_{-0.77}$    & $^{+0.60}_{-0.40}$ \rule{0pt}{2ex}  \\

\cite{Fatka2025}    & 5.021522      &  259      & +84               & 1.59      & 2.15\rule{0pt}{4ex}   \\
$3\sigma$-like unc. & $\pm0.000003$ &  \multicolumn{2}{c}{radius 8\degree}   & --        & --      \\

This work           &5.0215221      &  314     & +84                & 1.66      & 1.49\rule{0pt}{4ex}   \\
95\% bootstrap interval      &$^{+0.0000011}_{-0.0000007}$ &  \multicolumn{2}{c}{radius 5.4\degree}   & $^{+0.03}_{-0.07}$ & $^{+0.44}_{-0.29}$\\

\bottomrule
\end{tabular}

\par\vspace{0.5em}
\begin{minipage}{\textwidth}
\raggedright
\footnotesize
Note: For easier comparison, we express the asymmetric pole-coordinate uncertainties reported by  \citet{Popescu2025}, $(\Delta\lambda,\Delta\beta)=(\pm35^\circ,{}^{+1^\circ}_{-6^\circ})$, as an equivalent angular radius defined by the maximum angular separation from the nominal pole. The axis ratios from \cite{Popescu2025} are listed here in corrected form. In the original manuscript, the $b$ and $c$ axes were interchanged; the values shown here follow the $a \geq b \geq c$ convention used in this work (M. Popescu, private communication).
\end{minipage}
\end{table*}

\begin{table*}[tb]
\caption{\label{tab.results}Parameters of the best-fit convex model of asteroid (98943) Torifune and their bootstrap percentile intervals.}
\begin{tabular}{lrrrrr}
\toprule
             & $P_{sid}$ (h) &  $\lambda$ (\degree) & $\beta$ (\degree)& $a/b$    & $b/c$ \\
\midrule
Nominal             & 5.0215221 &  314     & +84.0     & 1.658     & 1.49\rule{0pt}{2ex}  \\

68\% boot-      &+0.0000005 &  +2      & +0.5      & +0.003    & +0.21\rule{0pt}{3ex}   \\
strap inter.           &-0.0000004 &  -30     & -1.5      & -0.042    & -0.14      \\

95\% boot-      &+0.0000011 &  +35     & +1.4      & +0.027   & +0.44\rule{0pt}{3ex}   \\
strap inter.   &-0.0000007 &  -54     & -3.9      & -0.066   & -0.29      \\

\bottomrule
\end{tabular}

\par\vspace{0.5em}
\begin{minipage}{\textwidth}
\raggedright
\footnotesize
Note: The table lists the nominal solution obtained from the inversion of the full dataset and the corresponding bootstrap percentile intervals. The 68\% interval is defined by the 16th and 84th percentiles, and the 95\% interval by the 2.5th and 97.5th percentiles. The pole coordinates $\lambda$ and $\beta$ are ecliptic longitude and latitude. Because the nominal pole lies close to the north ecliptic pole, the 68th and 95th percentiles of the angular-distance distribution with respect to the nominal pole direction, 2.1\degree\ and 5.4\degree, respectively, provide a more meaningful measure of pole uncertainty.
\end{minipage}

\end{table*}

\section{Implications for the Hayabusa2\# flyby}

The Hayabusa2\# flyby will provide a direct test of the pre-encounter convex model in a geometry that cannot be reproduced from Earth-based disk-integrated photometry. The most useful observables for this comparison will be the disk-resolved image sequence obtained by the Optical Navigation Camera Telescope (ONC-T), together with complementary thermal-infrared data from the Thermal Infrared Imager (TIR), during the final approach and closest-approach phase \citep{Hirabayashi2026}. These data will constrain the apparent orientation of the body, the projected limb, the global silhouette, and large-scale surface or thermal features not represented in a convex light-curve model. The mission light-curve data obtained before the flyby will provide an additional constraint on the rotational phase and on the position angle of the long axis as a function of time.


The present model makes two comparatively robust predictions. First, the spin state is tightly constrained: the sidereal period is unique, and the pole solutions form a compact cluster near the north ecliptic pole. Second, the equatorial elongation is relatively stable, as reflected in the well-constrained $a/b$ ratio. In contrast, the dimension parallel to the spin axis remains much less secure. The broad bootstrap distribution in $b/c$, together with the differences among published convex models, identifies the polar extent as the main shape parameter to be tested by the flyby. The resolved images and accompanying light curves should therefore be most informative for testing the projected global shape, apparent orientation, and $c$-axis extent, rather than for refining the already well-constrained mean rotation period.

Agreement between the observed ONC-T limb profiles and the predicted bootstrap silhouette envelope would support the conclusion that the available photometry reliably constrains the spin state and bulk convex shape of Torifune. Differences limited to local topography, concavities, roughness, or thermally distinct surface regions would be expected, because such features are outside the scope of convex light-curve inversion. A systematic mismatch in the global outline, apparent long-axis orientation, or polar extent would instead indicate that the Earth-based viewing geometries were insufficient to constrain the full three-dimensional figure.

The Torifune encounter will therefore serve as a useful benchmark for light-curve inversion of near-Earth asteroids. It will show which elements of the pre-flyby solution---period, pole direction, equatorial elongation, and polar flattening---are genuinely constrained by the current photometry, which are reliable enough for flyby planning, and which require additional information from resolved imaging, thermal-infrared observations, stellar occultations, radar, or non-convex modelling.

\section{Summary}
We derived an updated pre-flyby convex shape model of asteroid (98943) Torifune using previously obtained dense photometry data, newly published observations covering the early 2024 apparition, selected ATLAS sparse data extending the coverage to the late 2024 apparition, and our new observations from late 2025. The new 2025 observations sample viewing geometries over the northern hemisphere of the nominal convex model, which was previously less well covered (see Fig.~\ref{fig.observed_parts_all_2025}). Compared to the previous model, the present solution benefits from the extended temporal and geometrical coverage and from the weighting scheme designed to account for the heterogeneous quality and cadence of the input data.

A key methodological contribution of this work is the explicit treatment of data heterogeneity and model uncertainties. We introduced a per-measurement weighting scheme that accounts for the scatter of individual light curves, the partial correlation of closely spaced measurements, and phase-dependent variations in photometric quality. In addition, we estimated uncertainties by means of a block bootstrap in which complete light curves were resampled while preserving the dense/sparse composition of the dataset. To preserve the coverage of observing geometries, we further required that each bootstrap realisation include at least one light curve from each apparition represented in the original dataset. This approach provides uncertainty estimates not only for the sidereal period and pole direction, but also for the shape parameters, allowing us to quantify more robustly which aspects of the model are well constrained by the available photometry.

Our best-fit solution gives a sidereal period of $5.0215221$~h and a prograde spin axis at $(314\degree, +84\degree)$. The block bootstrap analysis shows that the rotational period and pole position are well constrained, while the convex shape is constrained more modestly, with $a/b$ better determined than $b/c$. We suspect that the weak constraint on $b/c$ arises from the limited range of viewing geometries, partly as a consequence of Torifune’s low orbital inclination ($\approx4.8\degree$). The anisotropic cumulative coverage shown in Fig.~\ref{fig.observed_parts_all} supports this interpretation.

The resulting model provides an improved prediction of the rotational state and global convex shape of Torifune ahead of the Hayabusa2\# flyby in July 2026. The flyby will offer a rare opportunity to distinguish between those components of a light-curve inversion solution that are robustly constrained by disk-integrated photometry and those that remain sensitive to limited observing geometry. In particular, the encounter should test whether the predicted pole orientation is consistent with the model, and whether global silhouette are confirmed by resolved imaging, while also constraining the poorly determined dimension parallel to the spin axis and revealing any major non-convex structure absent from the present model. Whether the agreement is close or imperfect, the Torifune encounter will therefore provide an important benchmark for assessing the reliability and limitations of convex light-curve inversion for near-Earth asteroids.

\section*{Data availability}
The new lightcurve photometry, derived model parameters, and associated machine-readable tables supporting this study are available through the Centre de Données astronomiques de Strasbourg (CDS) under catalogue identifier [The catalogue identifier will be provided upon acceptance].

\section*{Acknowledgements}
We are grateful to the reviewers for their careful reading of the manuscript and for their constructive comments, which helped improve the presentation and clarity of this work.

This work has been supported by the {\it Praemium Academiae} award from the Academy of Sciences of the Czech Republic, grant no. AP2401, and by the project RVO:67985815. 

This work uses data from the University of Hawaii's ATLAS project, funded through NASA grants NN12AR55G, 80NSSC18K0284, and 80NSSC18K1575, with contributions from the Queen's University Belfast, STScI, the South African Astronomical Observatory, and the Millennium Institute of Astrophysics, Chile.  

We used the publicly available code for the sidereal period search and the light curve inversion\footnote{\url{https://damit.cuni.cz/projects/damit/}} written by Josef \v{D}urech \citep[which is based on][]{Kaasalainen2001a, Kaasalainen2001b}. In this work we made extensive use of the programming language \verb|Python 3| \citep{python3} and its non-standard libraries \verb|NumPy| \citep{py_numpy}, \verb|SciPy| \citep{py_scipy}, \verb|Matplotlib| \citep{py_matplotlib}, and \verb|Astropy| \citep{py_astropy2013, py_astropy2018, py_astropy2022}. 

\section*{Declaration of generative AI and AI-assisted technologies in the manuscript preparation process}
The authors acknowledge the use of ChatGPT v5.4 and v5.5 (OpenAI) for coding assistance, data visualization, and language revision. All content was reviewed and validated by the authors, who take full responsibility for the content of the published article.

\appendix

\setcounter{table}{0}
\renewcommand{\thetable}{A.\arabic{table}}

\onecolumn
\begin{table*}[ht!]
\section{Summary of the light-curve dataset}

\caption{Summary of the photometric observations of asteroid (98943) Torifune used in the inversion.\label{tab.obs_all}} 
\centering
\begin{tabular}{lccl}
\hline\hline            
 Date span                & Observing runs & $\alpha$ ($^\circ$)  & Reference       \\
 \hline
2002-Jan-03 -- Jan-05     & 2         & 14 -- 12 & \citet{Fatka2025}                                    \\
2003-Nov-30 -- Dec-01     & 2         & 23 -- 24 & \citet{Fatka2025}       \\
2022-Jan-09 -- Mar-12     & 5         & 57 -- 93 & \citet{Fatka2025}                                \\
2022-Nov-29 -- Dec-29     & 2         & 47 -- 34 & \citet{Fatka2025}      \\
2023-Jan-22 -- Jan-22     & 1         & 21       & \citet{Fornasier2024} \\
2023-Feb-08 -- Feb-10     & 2         & 37 -- 39 & \citet{Fatka2025} \\
2024-Jan-06 -- Jan-15     & 2         & 22 -- 33 & \citet{Fornasier2024} \\
2025-Oct-23 -- Dec-20     & 7         & 21 -- 46 & this work \\
 \hline
  \multicolumn{4}{c}{\it ATLAS sparse data} \\
\hline
2023-Jan-13 -- Jan-23     & 5         & 24 -- 21 & ATLAS \\
2023-Dec-14 -- 2024-Jan-07& 4        & 12 -- 23 & ATLAS \\
2024-Nov-20 -- Dec-03    & 6         & 4 -- 13 & ATLAS \\
2025-Sep-23 -- Oct-28    & 16        & 61 -- 21 & ATLAS \\

\hline
\end{tabular}
\end{table*}

\section{Additional Figures}
\setcounter{figure}{0}
\renewcommand{\thefigure}{B.\arabic{figure}}

\begin{figure}[h!]
    \centering
    \includegraphics[width=0.6\hsize]{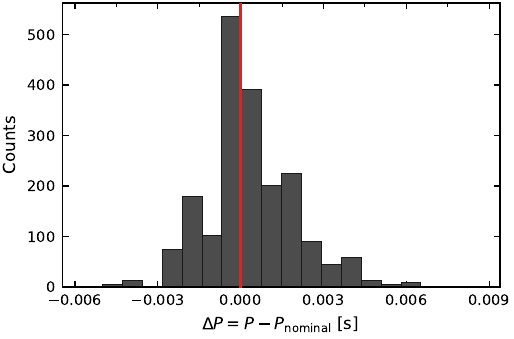}
    \caption{Distribution of best-fit sidereal periods from the 2000 block-bootstrap realisations. The red vertical line marks the nominal value, $P_{\mathrm{sid}} = 5.0215221$~h.}
     \label{fig.periods}
\end{figure}

\begin{figure}[h!]
   \centering
    \includegraphics[width=0.87\hsize]{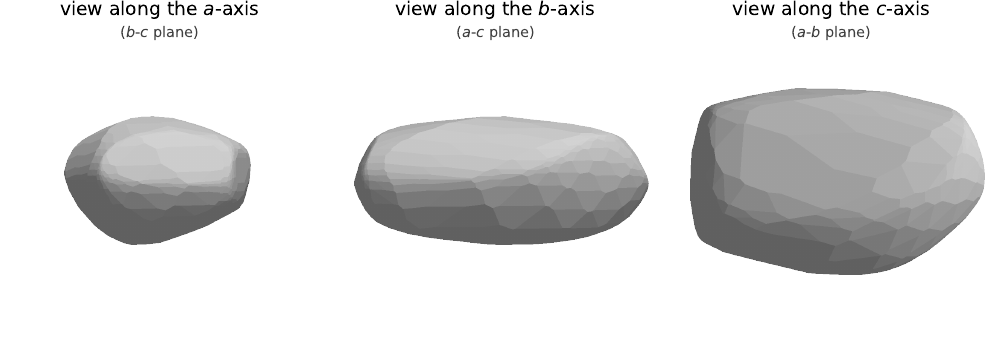}
    \caption{Best-fit convex shape model of asteroid (98943) Torifune shown at selected viewing aspects.}     \label{fig.best_shape}
\end{figure}

\begin{figure}[ht!]
   \centering
    \includegraphics[width=\textwidth,height=0.95\textheight,keepaspectratio]{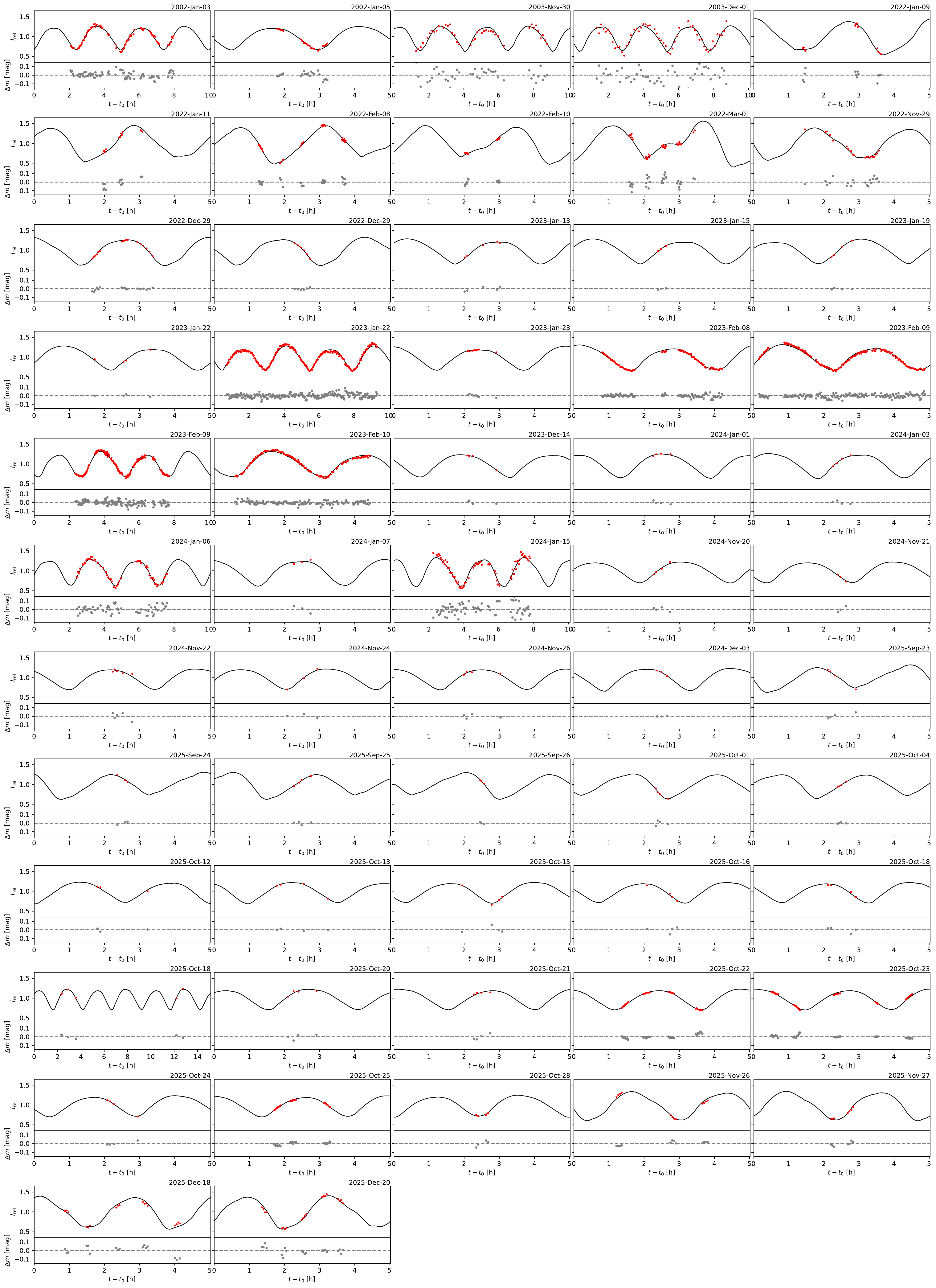}
    \caption{Comparison between the relative intensities of the observed light curves (red points) and the synthetic light curves of the best-fit model (solid curve). The corresponding residuals, expressed in magnitudes, are shown below each light curve as grey points.}    \label{fig.lc}
\end{figure}

Figure~\ref{fig.periods} provides additional diagnostics of the preferred solution. Figure~\ref{fig.best_shape} shows the best-fit convex shape model in three orthographic projections, whereas Fig.~\ref{fig.lc} compares the observed light curves with the synthetic light curves of the nominal model and shows the corresponding residuals.

\begin{figure}[h!]
   \centering
    \includegraphics[width=0.85\hsize]{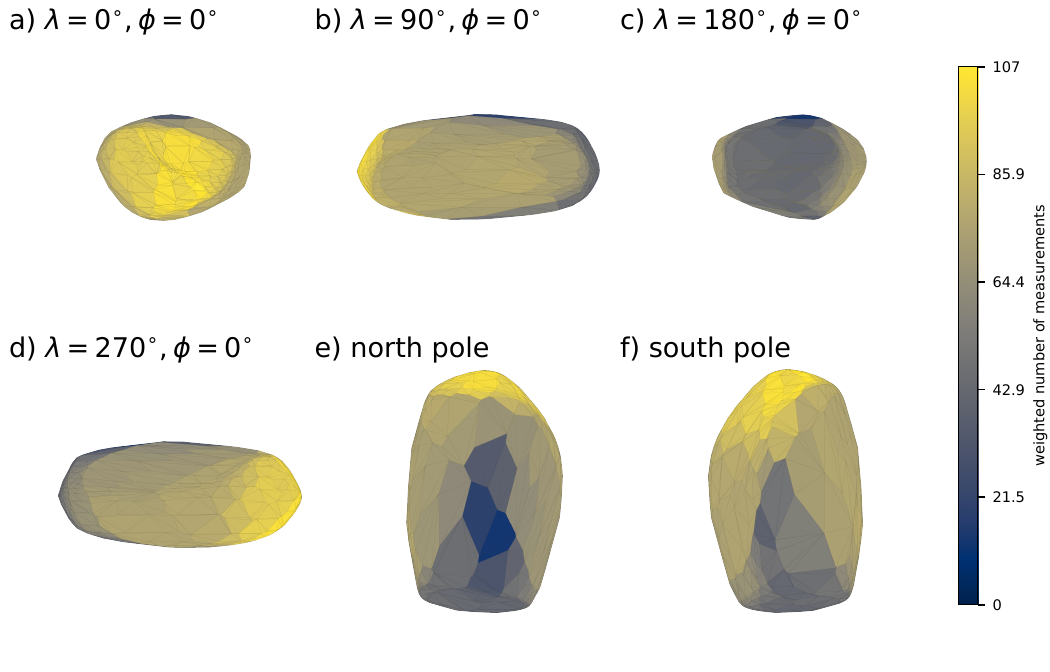}
    \caption{Same as Fig.~\ref{fig.observed_parts_all}, but for data obtained during the 2001-2002 apparition only.}     \label{fig.observed_parts_all_2001}
\end{figure}

\begin{figure}[h!]
   \centering
    \includegraphics[width=0.85\hsize]{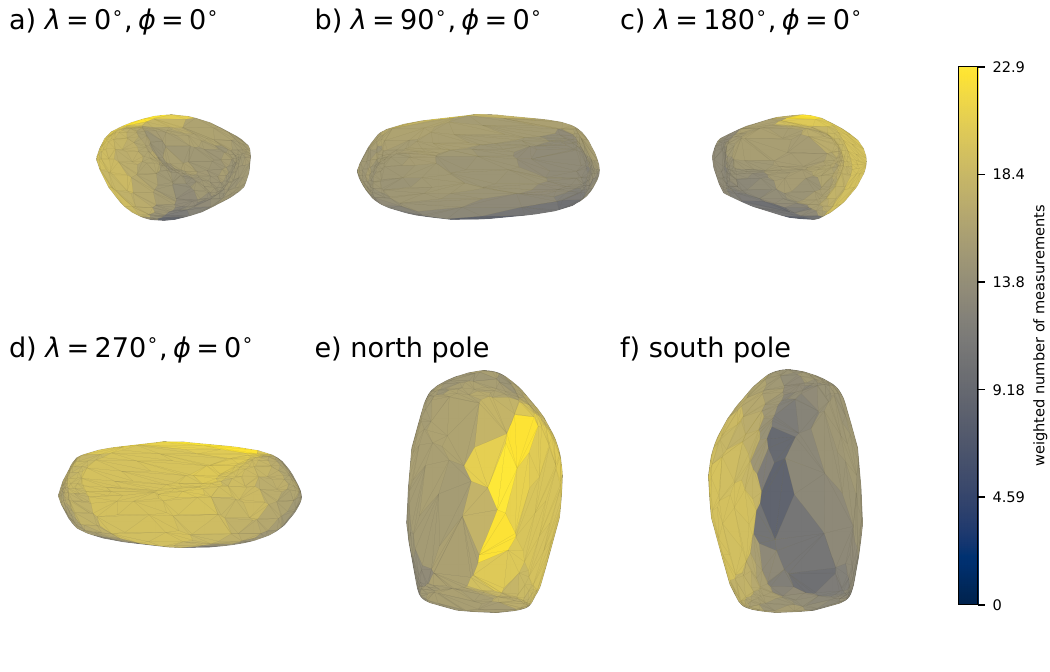}
    \caption{Same as Fig.~\ref{fig.observed_parts_all}, but for data obtained during the 2003-2004 apparition only.}     \label{fig.observed_parts_all_2003}
\end{figure}

\begin{figure}[h!]
   \centering
    \includegraphics[width=0.85\hsize]{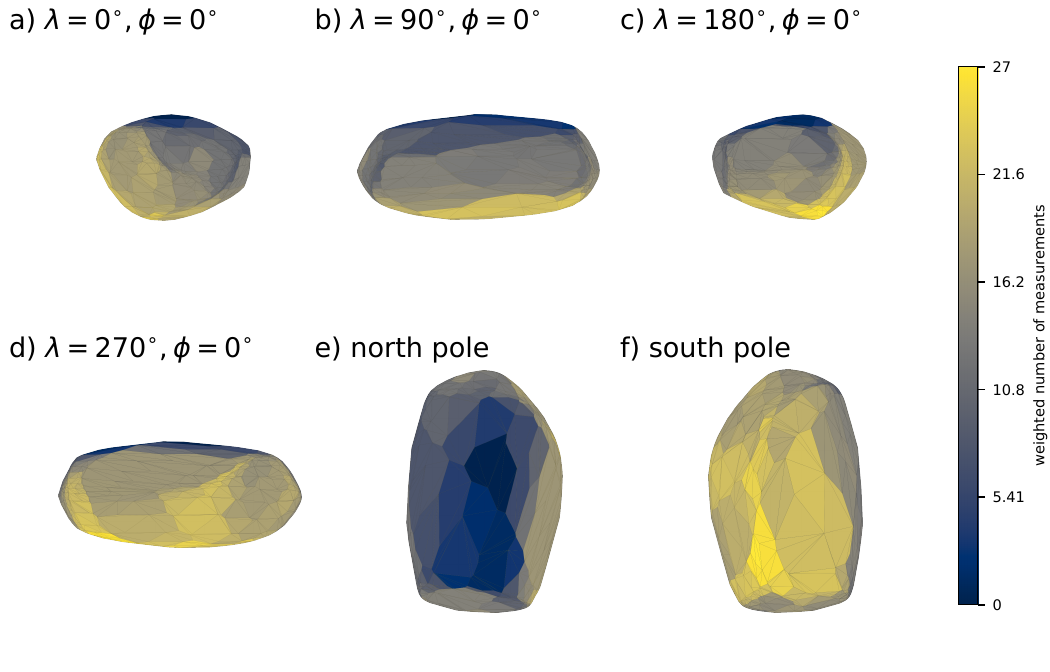}
    \caption{Same as Fig.~\ref{fig.observed_parts_all}, but for data obtained during the 2021-2022 apparition only.}     \label{fig.observed_parts_all_2021}
\end{figure}

\begin{figure}[h!]
   \centering
    \includegraphics[width=0.85\hsize]{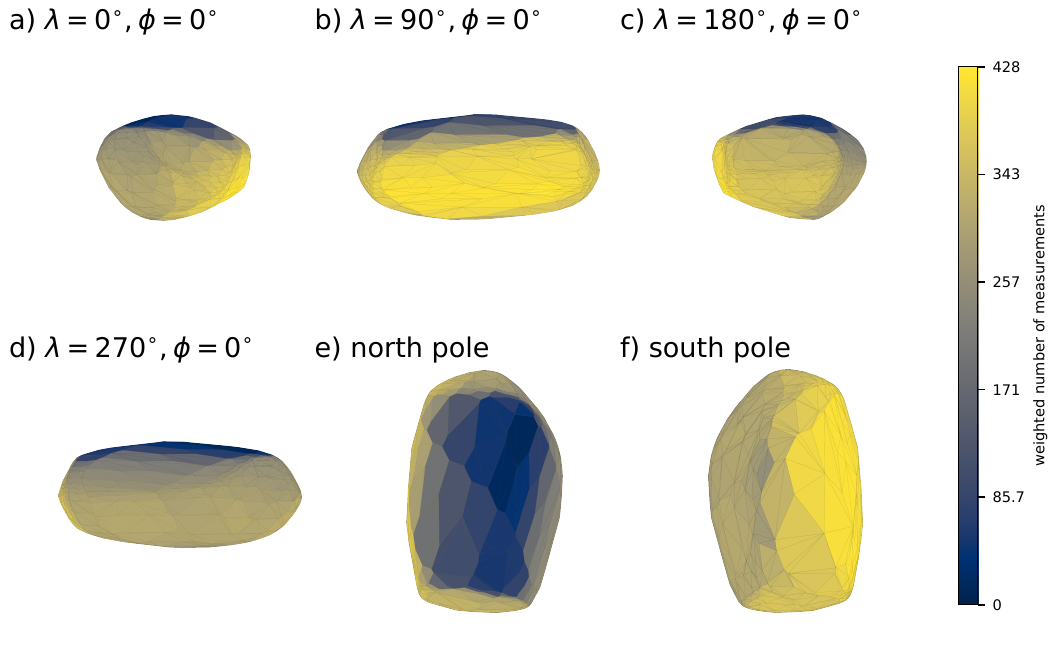}
    \caption{Same as Fig.~\ref{fig.observed_parts_all}, but for data obtained during the 2022-2023 apparition only.}     \label{fig.observed_parts_all_2022}
\end{figure}

\begin{figure}[h!]
   \centering
    \includegraphics[width=0.85\hsize]{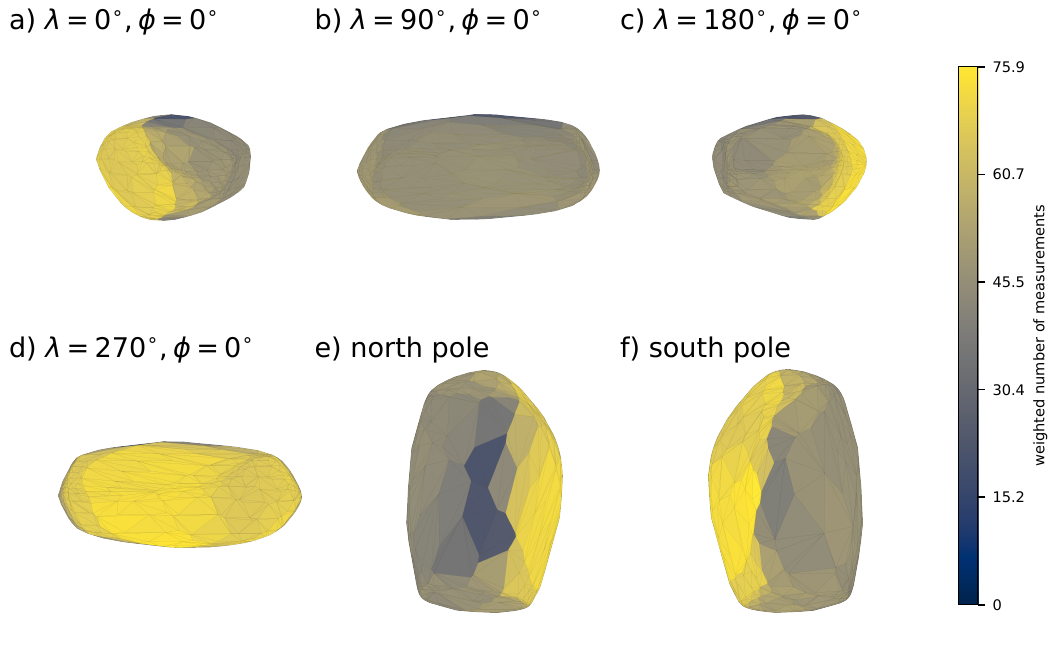}
    \caption{Same as Fig.~\ref{fig.observed_parts_all}, but for data obtained during the 2023-2024 apparition only.}     \label{fig.observed_parts_all_2023}
\end{figure}

\begin{figure}[h!]
   \centering
    \includegraphics[width=0.85\hsize]{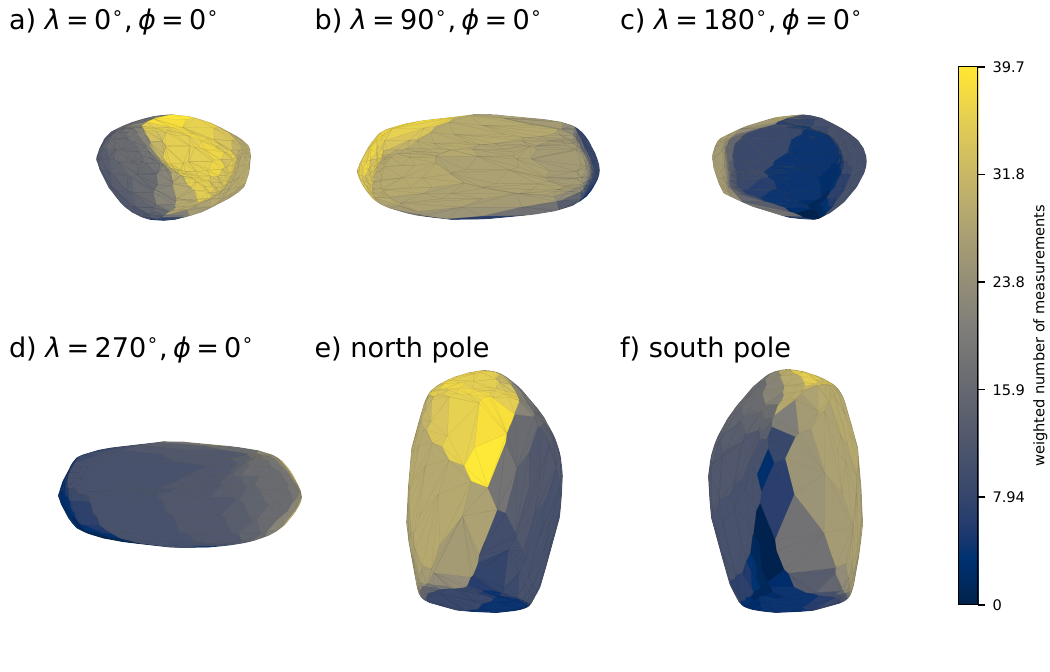}
    \caption{Same as Fig.~\ref{fig.observed_parts_all}, but for data obtained during the 2024-2025 apparition only.}     \label{fig.observed_parts_all_2024}
\end{figure}

\begin{figure}[h!]
   \centering
    \includegraphics[width=0.85\hsize]{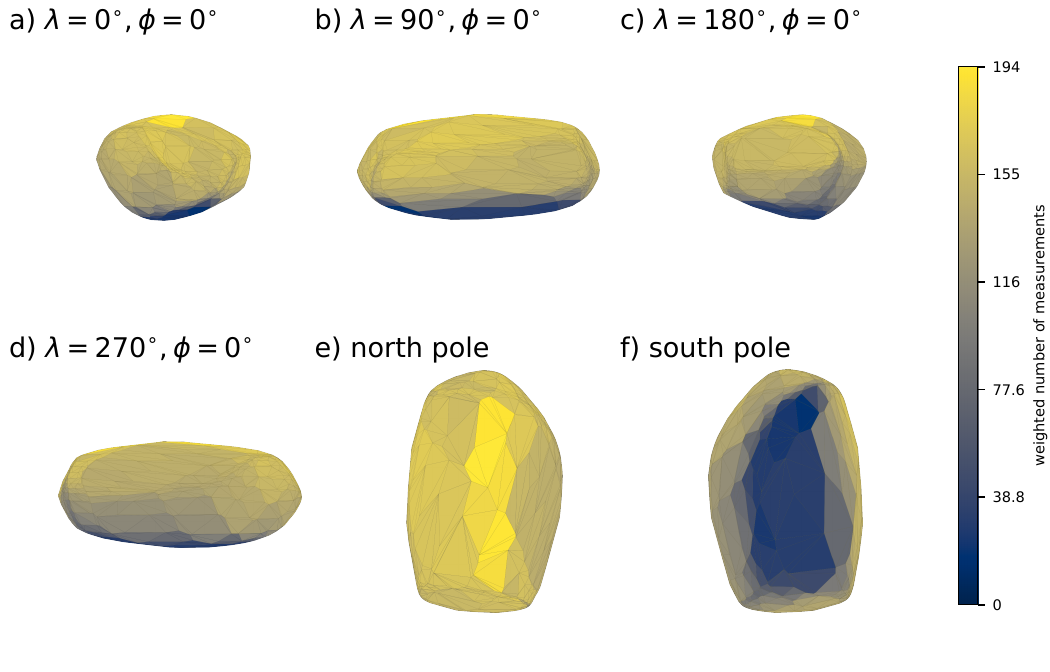}
    \caption{Same as Fig.~\ref{fig.observed_parts_all}, but for data obtained during the 2025-2026 apparition only.}     \label{fig.observed_parts_all_2025}
\end{figure}

\clearpage




\end{document}